# SIMBOL–X
# a formation flying mission for hard X-rays astrophysics

P. Ferrando[*a], A. Goldwurm[a], P. Laurent[a], O. Limousin[b], J. Martignac[b], F. Pinsard[b],
Y. Rio[b], J.P. Roques[c], O. Citterio[d], G. Pareschi[d], G. Tagliaferri[d], F. Fiore[e],
G. Malaguti[f], U. Briel[g], G. Hasinger[g], L. Strüder[g]
On behalf of the Simbol–X collaboration

[a] UMR 7164, APC & DSM/DAPNIA/Service d'Astrophysique, CEA/Saclay, 91191 Gif-sur-Yvette Cedex, France
[b] DSM/DAPNIA/Service d'Astrophysique, CEA/Saclay, 91191 Gif-sur-Yvette Cedex, France
[c] Centre d'Etude Spatiale des Rayonnements, BP 4346, 31028 Toulouse Cedex, France
[d] INAF, Osservatorio Astronomico di Brera, via E. Bianchi 46, 23807 Merate, Italy
[e] INAF, Osservatorio Astronomico di Roma, via E. Frascati 33, 00040 Monteporzio, Italy
[f] INAF, Istituto di Astrofisica Spaziale e Fisica Cosmica, Sezione di Bologna, via P. Gobetti 101, 40129 Bologna, Italy
[g] Max Planck Institut für Extraterrestrische Physik, 85748 Garching, Germany

**ABSTRACT**

SIMBOL-X is a hard X-ray mission, operating in the ~ 0.5–70 keV range, which is proposed by a consortium of European laboratories in response to the 2004 call for ideas of CNES for a scientific mission to be flown on a formation flying demonstrator. Relying on two spacecrafts in a formation flying configuration, SIMBOL–X uses for the first time a ~ 30 m focal length X–ray mirror to focus X–rays with energy above 10 keV, resulting in a two orders of magnitude improvement in angular resolution and sensitivity in the hard X-ray range with respect to non focusing techniques. The SIMBOL–X revolutionary instrumental capabilities will allow to elucidate outstanding questions in high energy astrophysics, related in particular to the physics of accretion onto compact objects, to the acceleration of particles to the highest energies, and to the nature of the Cosmic X–Ray background. The mission, which has gone through a thorough assessment study performed by CNES, is expected to start a competitive phase A in autumn 2005, leading to a flight decision at the end of 2006, for a launch in 2012. The mission science objectives, the current status of the instrumentation and mission design, as well as potential trade-offs are presented in this paper.

## 1. SCIENTIFIC AND PROGRAMMATIC CONTEXTS

The non thermal emissions in the X and gamma–ray domains reveal dynamical processes which involve very large transfers of matter and energy in a variety of astrophysical objects, at all scales from stars to clusters of galaxies. They therefore play a major role in the evolution of these objects as well as that of the Universe as a whole. Moreover, these processes occur in environments with extreme physical conditions in terms of gravity, pressure, density, or magnetic field, so that their study gives a unique opportunity to test the laws of physics well beyond the parameter space available in the laboratories. For these reasons the study of high energy phenomena is a major goal of modern astrophysics, currently addressed by a number of space missions, especially the large observatories, as XMM–Newton and Chandra in X–rays, and INTEGRAL in gamma–rays.

---

[*] Further author information: send correspondence to Philippe Ferrando, e-mail: philippe.ferrando@cea.fr

There is however a very large gap in spatial resolution and sensitivity between the X and gamma–ray domains. Below ~ 10 keV, astrophysics missions are using X–ray mirrors based on grazing incidence reflection properties. This allows to achieve an extremely good angular resolution, down to 0.5 arcsec for Chandra, and a good signal to noise thanks to the focusing of the X–rays onto a small detector surface. This technique has so far been limited to energies below ~ 10 keV because of the maximum focal length that can fit in a single spacecraft. Hard X–ray and gamma–ray imaging instruments are thus using a different technique, that of coded masks. This non-focusing technique has intrinsically a much lower signal to noise ratio than that of a focusing instrument, and does not allow to reach angular resolutions better than a few arc minutes. In addition to the difference in angular resolution, there is also roughly two orders of magnitude of difference in point source sensitivity between X–ray and gamma–ray telescopes.

This transition of techniques unfortunately happens roughly at the energy above which the identification of a non-thermal component is unambiguous with respect to thermal emission. Considered from the low energy side, this obviously strongly limits the interpretation of the high quality X–ray measurements, and particularly that related to the acceleration of particles. Considered from the high energy side, this renders impossible the mapping of the gamma–ray emission of extended sources to the scales needed to understand the emission mechanisms by comparing with lower energy data, and this limits the studies to very bright sources only.

A clear requirement for future high energy astrophysics missions is now to bridge this gap of sensitivity, by offering an instrumentation in the hard X–ray range with a sensitivity and angular resolution similar to that of the current X–ray telescopes. In order to do this, a hard X–ray focusing optics is needed. Such an optics can readily be implemented by a simple extension of the current X–ray mirror technology to long focal lengths, which is the basic concept of the Simbol–X mission.

At the same time, there is now the emerging technical possibility to design missions using multiple spacecrafts flying in a constrained formation. At the beginning of 2004, the CNES space agency has issued a call for proposal of a scientific payload to be flown on a formation flight demonstrator mission to be launched at the beginning of the next decade. The Simbol–X mission has been proposed to CNES in this context, and has been selected for further mission implementation studies.

Simbol–X is a hard X–ray pointed telescope, based on a very long focal length optics (30 m in nominal design) which will extend the focusing technics to energies up to at least ~ 70 keV, offering a gain of roughly two orders of magnitude in sensitivity and angular resolution compared to the current instruments above 10 keV. In addition to this fundamental breakthrough in the hard X–rays, Simbol–X will have a low energy threshold around ~ 0.5–1 keV, which will allow to fully cover the transition from thermal to non-thermal emissions, as well as the Iron line region, two important characteristics for the study of the highly variable accreting sources which are prime scientific targets of this mission.

Simbol–X has undergone a thorough assessment study (called "phase 0") performed by a dedicated CNES engineering team (PASO) in the beginning of 2005. This study, which has covered all the aspects of the mission from payload to ground segment, has demonstrated that the Simbol–X payload and mission implementation fit well within the constraints of the call for proposal issued by CNES.

An extensive description of Simbol–X has already been given by Ferrando *et al.*[1], with a first payload design and a possible mission implementation (also issued from an earlier CNES study). The basics of the mission remain the same, and will be only concisely repeated here. We will concentrate more on the important changes since that paper, both on the payload and on the mission implementation.

## 2. SCIENTIFIC OBJECTIVES

Offering "soft X–ray"-like angular resolution and sensitivity in the hard X–ray range, SIMBOL–X will provide a dramatic improvement for investigating a number of fundamental problems in astrophysics, which have their essential signatures in non thermal emissions. The recent measurements made by the IBIS/ISGRI instrument onboard INTEGRAL have unveiled even further the richness of the hard X–ray domain, in particular by discovering numerous highly absorbed sources, the nature of which remains a puzzle[2]. The list of detailed topics and subjects that Simbol–X can address is very large, and cannot be detailed here. We simply mention below the three main scientific questions in which Simbol–X will make a breakthrough. More information and discussions, as well as detailed simulations of observations in the nominal design, can be found in paper I.

The first scientific question that Simbol–X will address is the dynamics of matter around compact objects, and in particular Black Holes. By providing the most extreme gravitational conditions under which matter can be presently observed, the Black Hole environment is a unique laboratory to test laws of physics. If it is generally admitted that the accretion of matter onto these objects occurs via a very dynamic accretion disk, the way that this disk evolves, the mechanisms that form the often observed relativistic jets, as well as the origin of the high energy radiation remain unclear. Finding the origin of the observed emission, between e.g. synchrotron in a jet, Compton from a hot corona, or reflection from the disk, and identifying the geometry of the system necessitates accurate measurements of the non thermal spectrum together with that of the Iron line shape. Simbol–X will allow such measurements on Black Holes of all masses in up to now uncharted regimes or locations.

For Galactic Black Holes of stellar mass in binary systems, these include spectral measurements in all states of accretion, and for the first time in the elusive quiescent state. A very large sample of binary systems in nearby Galaxies, up to ~ 3 Mpc, will also be measured for the first time in hard X–rays, providing absolute luminosities for these systems, and allowing population studies for other galaxies than ours. Going to higher mass systems, Simbol–X will be able to observe and resolve for the first time Ultra Luminous X–ray sources above 10 keV. This is a key ingredient to understand the nature of these sources, which are suspected to harbour an intermediate mass black hole, but which might be of a more classical nature, like e.g. QSOs. At the end of the black hole mass scale, Simbol–X will provide very detailed spectra for a very large sample of AGNs on their dynamic time scale (a few ks). As for binary systems, this will allow to disentangle the different possible emitting components, to understand the geometry of the systems, and measure the Black Hole spin via the Fe line shape. Moreover, Simbol–X will allow a true population study for the high energy emission of obscured AGNs, which are playing a key role in X–ray background synthesis models. Today, only a handful of them are known. Statistically, Simbol–X will increase by a factor of more than 100 the volume of Universe in which AGNs can be studied at high energy. Finally, but not least, Simbol–X will give a definitive measurement of the high energy emission of the supermassive black hole at the centre of our Galaxy, SgrA*, which is still unknown despite deep INTEGRAL observations[3]. The understanding of this object and its relations with its surroundings, as well as the origin of the puzzling high energy diffuse emission in the centre of the Milky Way, is crucial for understanding the much farther away galactic cores.

The second main scientific question that Simbol–X will address is that of the acceleration of particles in the Universe, in relation with the still unsolved problem of the origin of the cosmic-rays, particularly of the acceleration sites for the highest energies.

For that purpose, Simbol–X will observe acceleration in supernova remnants, for which the non thermal emission arises from electrons of tens of TeV energies. Measuring the synchrotron spectrum into the hard X–rays will allow to determine the maximum energy of the accelerated electrons, and will give unique insights in the mechanism responsible for the limitation in energy. In addition, Simbol–X will also map the supernova remnants, and correlate these data with radio, and now available TeV gamma–rays data from HESS (e.g. Aharonian *et al.*[4]). Putting these together will unambiguously allow to disentangle the different components, in particular the expected but not yet firmly identified hadronic one. Simbol–X will also observe other known sites of particle acceleration, like in particular extended X–ray jets of AGNs, as e.g. those of Cen A and Pictor A, which above 10 keV cannot be separated today from the bright emission of the accreting object. This is crucial to understand the emission mechanisms at work in these jets (synchrotron, inverse Compton), and to measure the maximum energy of the accelerated electrons.

The third major scientific question that will be addressed with Simbol–X is that of the structure of the Universe and of its evolution. Key observables are clusters of Galaxies, for which Simbol–X will definitely measure the controversial presence of a non thermal component, and the diffuse Cosmic X–Ray Background (CXRB). The CXRB is the most direct probe of accretion activity onto super-massive black holes, and determining its origin gives major constraints for the formation and evolution of structures in the Universe. If the major fraction of the CXRB has been resolved into discrete sources below ~ 7 keV, only half of it has been accounted for in the 7–10 keV band, and essentially nothing is directly known above 10 keV, and in particular in the 30–50 keV band in which the CXRB spectrum is peaking. Observational and theoretical evidences based on soft X–Rays deep fields have led to the conclusion that a significant fraction of the CXRB is due to obscured Compton thick sources. However, such a conclusion remains unconstrained by the available observations, and a major breakthrough is needed to validate the CXRB synthesis models and to assess the values of their parameters, such as the space density and evolution of these obscured sources. With about a 1 µCrab flux sensitivity in the 20–40 keV band, Simbol–X will have the necessary power to detect these Compton thick sources and resolve between 35 and 65 % of the CXRB in the energy range where it peaks, thus providing a true quantum leap forward for the understanding of this fundamental component of the Universe.

## 3. MISSION CONCEPT AND CHARACTERISTICS

Simbol–X is basically built using a classical Wolter I optics focusing X–rays onto a focal plane detector system. The gain in maximum energy is achieved by having a long focal length, of 30 metres in the nominal design studied so far, i.e. 4 times that of XMM–Newton mirrors. Since this cannot fit in a single spacecraft, the mirror and detectors will be flown on two separate spacecrafts in a formation flying configuration, as explained in paper I and sketched on Figure 1.

Simbol–X is a pointed telescope, which is nominally required to be able perform very long uninterrupted observations (100 ks or more) on the same target. The necessity to have a stable image quality, as well as to keep the full field of view inside the detector area, dictate the requirements on the formation flying stability. The constraint is that the distance between the two spacecrafts (along the telescope axis) must be kept at the focal length value within about ± 10 cm, whereas the intersection of the telescope axis must be at the centre of the focal plane within about ± 20 arcsec. It is also required that the arrival direction of each detected photon can be reconstructed to ± 1.5 arcsec, which imposes a knowledge (monitoring) of the relative positions of the two spacecrafts to that level of accuracy.

Compared to the design and performances described in paper I, several evolutions have taken place or will be considered in the phase A study. The first is the possibility of improving the Simbol–X performances by modification of the optics from the nominal design. The second is a careful investigation of the baffling issue against the diffuse X–ray sky background. The last one is on the spacecrafts design and mission implementation. They are described with some details below, while the reader is referred to paper I for a more extensive description of the "basics" of the payload and mission.

### 3.1 Optics
The Simbol-X optics is a Wolter I nested shells mirror. These shells will be made following the Nickel electroforming replication method[5], for which the Brera Observatory and its associated industrial partners have acquired a large experience through the building of the BeppoSAX, SWIFT, and XMM–Newton mirrors. The long focal length coupled to the requirement to have a large filling factor lead to a total number of shells of about 100. Compared to the XMM–Newton mirror case, the thickness to diameter ratio for the shells will be reduced so that the mass of the mirror stay within the mission constraints. At the current level of overall design of the mission, the mass mirror allocation is about 300 kg. With such a mass, and for this number of shells, the angular resolution will be no worse than 30 arcsec (FWHM) as experimentally proven[6]. The goal for angular resolution is to reach 15 arcsec, which might necessitate a trade-off with the effective area. This will be considered in the phase A study.

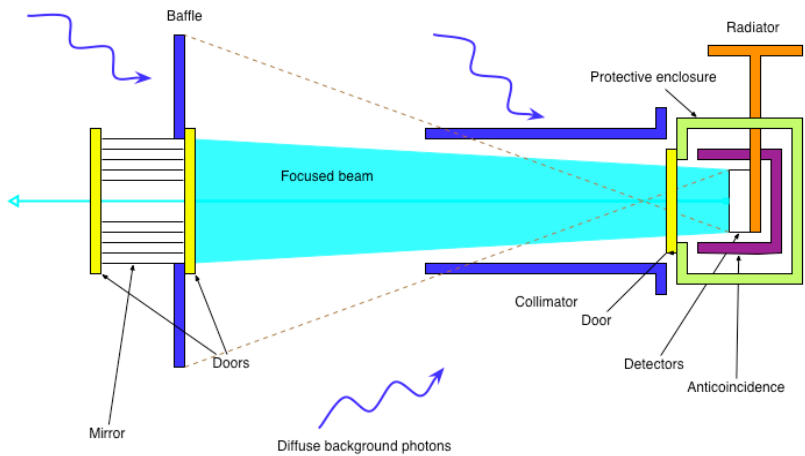

Figure 1 : Sketch of the SIMBOL–X configuration. The mirror, on the left, is carried by one spacecraft, while the detector system and the collimator, on the right, are carried by a second spacecraft. The figure emphasizes the issue of the diffuse sky background, discussed in Section 3.4, and is therefore not to scale. Typical sizes for a 30 m focal length (distance between the two spacecrafts) are 3 m for the baffle diameter, 1.5 m for the collimator length and 10 cm for its diametre.

The nominal coating considered in paper I is made of a conventional single layer of Pt. The optics properties then entirely rely on the total reflection phenomenon, which is characterized by a very large reflectivity at grazing angles until a critical angle beyond which the reflectivity falls rapidly down to almost zero. Beyond 10 keV the useful angles for reflection become very small, determining, as a consequence, a strong limitation of the available collecting area for optics with usual focal lengths ($\leq 10$ m). This problem is solved with the possibility of having very long focal lengths as offered by the formation flight technology. Large collecting areas can then be obtained at several tens keV with a conventional single layer coating. This is the baseline given in paper I, with a 30 m focal length. The effective area for this single layer baseline is plotted as the dashed curve in Figure 2a.

Besides this baseline design, we are now considering replacing the single Pt layer coating by a multi-layer Pt/C coating, while still remaining compliant with the mission constraints. The description and optimization of the multi-layer reflectors can be found in Cotroneo & Pareschi[7]. There are two advantages in using a multi-layer coating. The first one is to increase very significantly the energy range up to 80 keV and above. This will allow to effectively cover the region of the $^{44}$Sc lines (68 keV and 78 keV) resulting from the decay of $^{44}$Ti, an objective of prime importance for the astrophysics of supernovae explosion. The second one is to increase the field of view, provided that the focal length is decreased. This would obviously reduce the time necessary to map extended objects, and increase the statistics of the discovery of highly obscured sources per observation. Examples of what can be achieved are shown in Figure 2a and 2b, for two extreme cases.

In Fig. 2a a simple replacement of the single layer Pt coating by a multi-layer has been simulated, while keeping the same geometry and number of shells. There is a very important gain of effective area, which is still over 30 cm$^2$ beyond 80 keV. There is also a slight gain in the field of view, which increases from 6 to 7 arcmin (diameter for 50 % vignetting) with the multi-layer option.

In Fig. 2b, two different geometric configurations are considered; one is the same as in Fig. 2a (30 m focal length, 60 cm outer diameter), while the other one assumes for the outer diameter the maximum value possible to build with existing facilities, as well as a shorter focal length of 20 m. This last option has the advantage of increasing very significantly the field of view up to 12 arcmin, as well as the effective area below ~ 25 keV, but the drawback of having much less effective area above 30 keV, which falls essentially to zero at the Pt K edge at 78.4 keV.

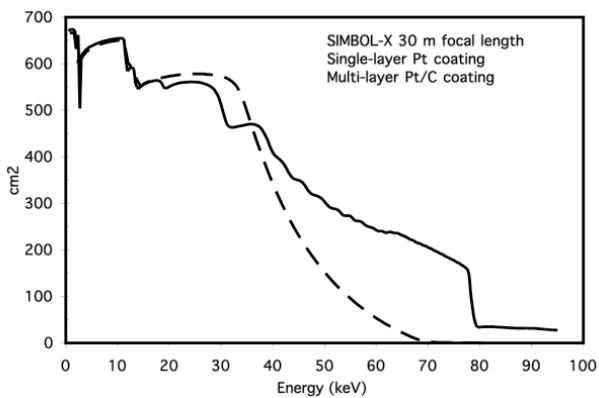 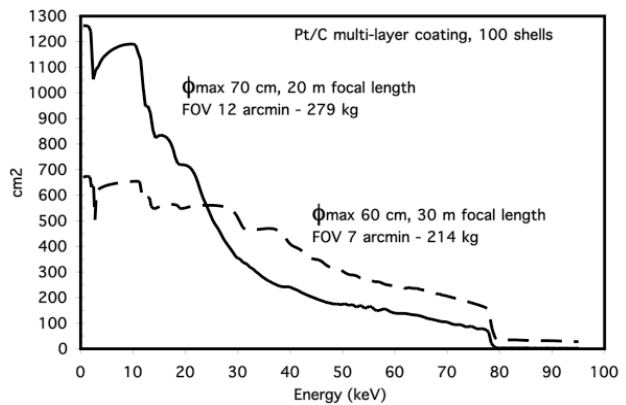

Figure 2a : Mirror effective area of the 60 cm diametre mirror, with a single-layer Pt coating (dashed line), and with a multi-layer coating (full line).

Figure 2b : Mirror effective area for two extreme cases. Both are assuming a multi-layer coating. The dashed line is the same as in Fig. 2a (note the difference of scale). The full line is assuming a larger outer diametre, and a shorter focal length.

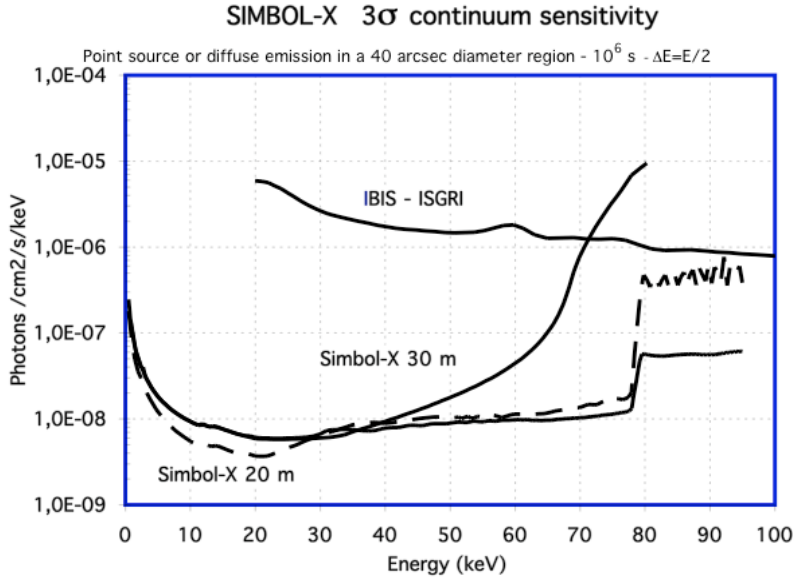

Figure 3 : continuum sensitivity of Simbol–X, for the three different options of optics corresponding to Fig. 2. The sensitivity is calculated with the same internal background assumptions as in paper I, i.e. about $7\ 10^{-5}$ cts/cm$^2$/keV. The sensitivity is calculated for 1 Ms of observation, for a better comparison with INTEGRAL, and $\Delta E = E/2$. The two Simbol–X full line curves correspond to the 30 m focal length with single Pt layer coating and multi-layer coating respectively. The dashed curve is the for the 20 m focal length case, with multi-layer coating.

These two extreme cases illustrate the range of performances that can be expected for the final design. This will be optimized for the best trade-off between effective area and field of view with the respect to the different science objectives, taking also into account the formation flying constraints.

On the other hand Figure 3 displays the Simbol–X sensitivity to the detection of point sources, for the three optics cases described above, assuming a 40 arcsec extraction region which contains 90 % of the source photons. It is interesting to see that the two multi-layer options have in fact a very similar sensitivity up to the Pt K edge, despite their different effective area. The reason is that, compared to the 30 m focal length case, the smallest effective area of the 20 m focal length case is compensated by the fact that the background in the extraction zone is also smaller because of the smaller plate scale. On this figure has also been plotted the IBIS/ISGRI sensitivity to point sources, for the same observation parameters. One can readily see that more than two orders of magnitude are gained by Simbol–X, up to about 50 keV in the single layer Pt coating case, and about 80 keV if a multi-layer coating is assumed.

### 3.2 Thermics and low energy threshold

Even if Simbol–X is designed for non-thermal hard X–ray astrophysics goals, it is desirable to have the lowest energy threshold possible. This is in particular the case for all variable objects for which a measurement of the thermal component contemporary to the non-thermal one is necessary, in order to gain access to parameters like for example the absorption column or the disk temperature in binary systems. In order to do so, there should be the less possible material in the path of the X–rays.

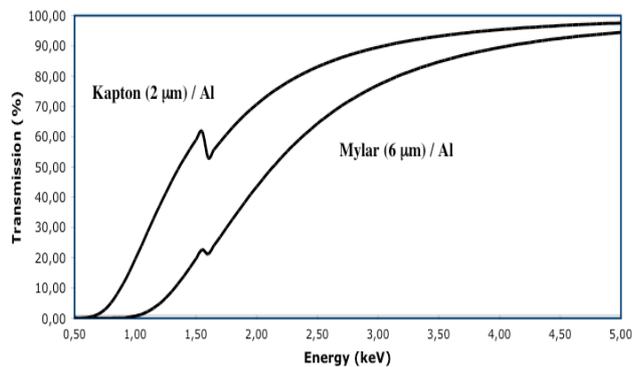

Figure 4 : Overall transmission of filter foils, in two different cases. The Aluminum thickness is 1000 Angströms

Ideally this means in particular that the mirror shells should be fully open to the outer space, like they are on Chandra or XMM–Newton. This situation is however very demanding for the thermal control and stresses on the mirror module, especially since in the case of the formation flying configuration both sides of the mirror are open to space. In order to minimize these thermal loads, it is anticipated to cover each side of the mirror by a thin aluminized foil. On Figure 4 are shown for example the transmissions resulting from the coverage of each mirror side by a single aluminized foil supported by different material. One case is for 6 μm thick Mylar, the other case for or a 2 μm thick Kapton. For a threshold definition at 20 % of transmission, this gives an energy threshold of ~ 1.5 and 1.0 keV respectively. Such thicknesses are in fact rather large, and foils with sub-micron support can be manufactured and have been flown[8]; minimizing the thickness given the mechanical constraints will be performed in phase A, but it is already anticipated that the low energy threshold will be between 0.5 and 1 keV at most. Such foils would moreover have the advantage of providing the blocking of the optical light.

### 3.3 Telescope characteristics

The table below summarizes the characteristics of Simbol–X discussed in the previous sections. The left column is the baseline design similar to that of paper I. The right column indicates the areas where an improvement can be made by the use of a multi-layer coating, which will be studied in phase A. In addition to the parameters already discussed, the requirement on the energy resolution for the hard X–rays will depend on the option. The only lines necessitating an important energy resolution are those linked to the $^{44}$Ti decay, which are just above the limit of detection of current instruments like INTEGRAL. Simbol–X will not do better in the single layer case, so that a relatively moderate energy resolution is sufficient. In the multi-layer case, Simbol–X has definitely the power to measure the lines profiles, at least in the known case of CasA, which explains why the requirement on energy resolution is more stringent.

| Parameter | Baseline (Pt single layer) | Changes with multi-layer |
|---|---|---|
| Energy range | < 1.0 keV – 70 keV | < 1.0 – 100 keV |
| Energy resolution | ~ 130 eV @ 6 keV <br> ~ 2 keV @ 60 keV | – <br> 1 keV @ 60 keV |
| Angular resolution | < 30 arcsec HEW, goal 15 arcsec | – |
| Localisation | < 3 arcsec | – |
| Field of View (50 % vignetting) | 6 arcmin | 7 – 12 arcmin |
| Effective area | 550 cm$^2$ below 35 keV <br> 2 cm$^2$ at 70 keV | 550 – 1100 cm$^2$ below 25 keV <br> 200 cm$^2$ at 70 keV |
| Sensitivity (3 σ, 1 Ms, dE/E = E/2) | 10$^{-8}$ ph/cm$^2$/s/keV for E < 40 keV | 10$^{-8}$ ph/cm$^2$/s/keV up to 80 keV |

### 3.4 Baffling the sky background

As there is no telescope tube between the mirror and the focal plane unit in a formation flight configuration, there is a need to protect the detectors from seeing X–rays coming from outside of the mirror, which would otherwise generate an unwanted background. When integrated over even a modest solid angle, the Cosmic diffuse X–ray background has indeed a level much larger than what is required to reach the desired sensitivity. As a matter of fact, in the case of a 2π opening of the detectors on the sky, the sensitivity would be reduced by two orders of magnitude, i.e. the gain in sensitivity brought by the focusing would be entirely lost. This is discussed also in details by Malaguti *et al.*[9]. For Simbol–X, we have decided to entirely baffle the sky, so that only X–rays focused by the mirror reaches the focal plane unit. In order to do that, we combine a "sky baffle" placed on the mirror spacecraft around the mirror, and a collimator placed on top of the focal plane, as illustrated in Fig. 1. Both the sky baffle and the collimator will be graded shields to stop the highest X–ray background energies and to avoid unwanted fluorescence lines. The Figure 5 displays the length of the collimator needed to entirely shield the diffuse sky, as a function of the size of sky baffle on the mirror spacecraft. As expected, the largest is the sky baffle, the shortest needs to be the collimator. As we will see, we can accommodate a 3 m diameter sky baffle, so that a ~ 1.5 m collimator is sufficient. The other curve on Figure 5 is the total mass of the baffling system, i.e. that of the collimator added to that of the sky baffle. The mass has been calculated by putting the amount of material needed to decrease the transmitted diffuse background flux below the internal detector background level. Since the solid angle covered by the sky baffle is quite small, the baffle does not need to be as opaque as the collimator which has to suppress a much larger integrated background flux. The mass of the sky baffle remains relatively small even for very large diameters. Overall, we have calculated that the mass of the shielding material is less than 20 kg for the 3 m diameter case. In Figure 5, we have added to this stopping material mass a conservative estimate for an additional structural mass to hold the shielding, leading to an overall mass of about 35 kg for the complete system.

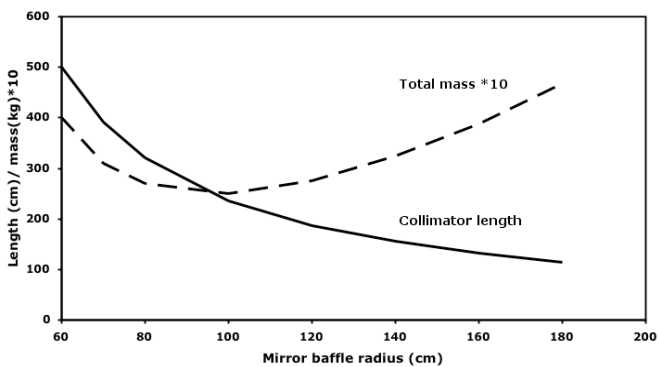

Figure 5 : Length of the collimator, and mass of the total baffling system, as a function of the sky baffle radius.

### 3.5 Focal plane unit

The focal plane is essentially similar to that described in paper I, to which the reader is referred for a detailed description. We simply recall for the sake of completeness the main characteristics of the detector system. The focal plane detector system thus consists in the combination of a low energy detector made of a DEPFET-SDD Active Pixel Sensor, efficient up to ~ 20 keV, on top of a mosaic of high energy Cd(Zn)Te pixelized detector. The two cameras will be surrounded by an active anticoincidence shield. Both detectors are spectro-imagers, and will have a pixel size of ~ 500 µm, which will allow a sufficient over sampling of the mirror point spread function. The two detectors have very fast reading capabilities, and it is envisioned to also use the SDD detector signal in the anticoincidence scheme of the Cd(Zn)Te detector which will then be entirely surrounded by and active anticoincidence shielding.

Detector teams at MPE (DEPFET-SDD low energy detector) and at CEA/DAPNIA (Cd(Zn)Te high energy detector) are currently concentrating on manufacturing and testing prototypes of the low and high energy detection units. On one hand, an 8 × 8 APS prototype based on the DEPFET–SDD combination with 1 × 1 mm$^2$ pixels has been successfully produced. First tests of this device show very promising results. The results of the similar XEUS DEPMOSFET matrix can be found in Treis *et al.*[10]. On the other hand, an 8 × 8 CdZnTe prototype, 1 mm pitch, equipped with its dedicated full custom read-out ASIC IDeF-X V1.0 has also been successfully implemented. The most recent detector characteristics and ASIC performances will be published respectively in references [11] and [12].

## 4. MISSION STUDY

As mentioned in Section 1, the Simbol–X mission has been studied by a dedicated engineering team of CNES, the PASO. The goal of this study, so called phase 0, was to fully assess the feasibility of the mission, and its cost, as an input in the selection process of the scientific mission to be flown as a formation flying demonstrator. This study has covered all aspects of the mission and its implementation, going well beyond the initial study made in 2003 to which we referred to in paper I. We simply mention briefly here a few points that have come out from this study, especially those in relation with the scientific operations.

The first point is about the orbit. Given the requirements to observe beyond 75,000 km of altitude to minimize the radiation level, and the envisioned launcher, the most favorable orbit that came out for Simbol–X is a high elliptical orbit, with a perigee of 44,000 km at launch, and an apogee at 253,000 km. This orbit has a period of 7 days, and allows continuous observations for most of the time, and may be all the time if the level of radiation actually encountered close to the (high altitude) perigee is small enough. Moreover, such high altitudes fully minimize the gaz consumption needed for keeping the formation, since the gravity gradient between the two spacecrafts is smaller than the radiation pressure. The telescope axis orientation is constrained to be perpendicular to the Sun to Simbol–X line within ± 20 degrees, which provides a full sky visibility in about 4.5 months.

The orbit can be essentially fully covered by two ground stations, separated by 12 hours. It is however not necessary to have a permanent link with Simbol–X, since the average level of telemetry will be relatively low (about 10 Gbits per week for an average observation plan), and since the system has to be autonomous for its safety in the formation flight configuration. The baseline scheme is to have Simbol–X autonomous during observations, with data recorded onboard. The contact with Simbol–X is made nominally only for a change of target, which is conducted and followed in real time by the ground. At the same time, scientific data of the previous observation can be downloaded.

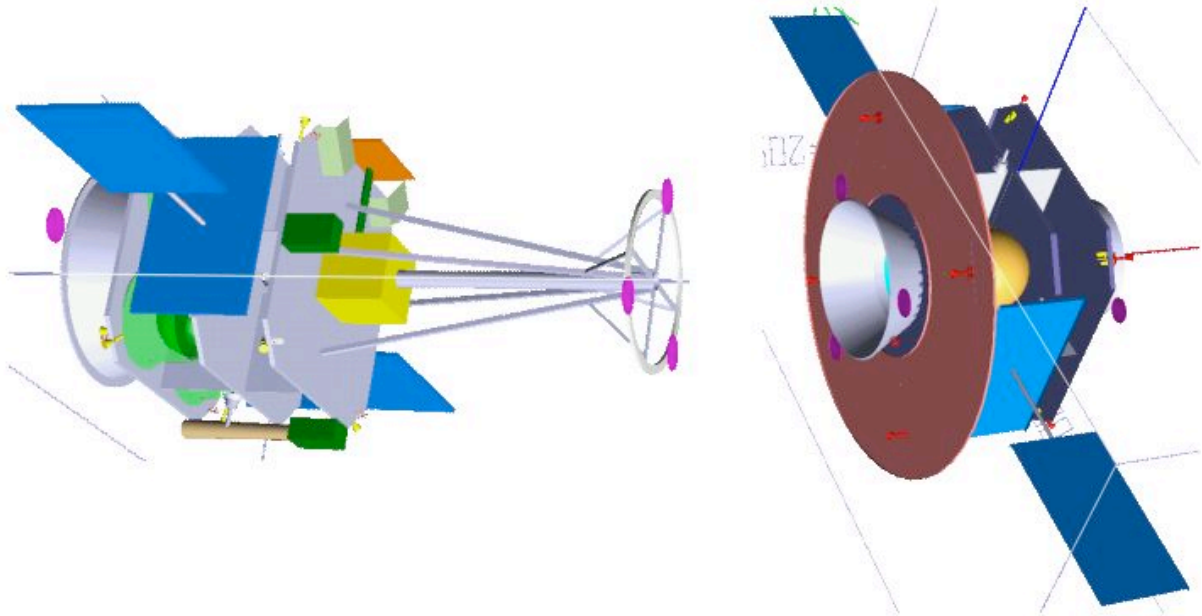

Figure 6 : Detector (left) and mirror (right) spacecrafts. The scale is not the same for the two spacecrafts. On the detector spacecraft, one can see the focal plane unit with a ~ 1.5 m collimator on top. On the mirror spacecraft, the prominent circular structure is the 3 m diameter sky baffle.

In view of the various scientific domains that will be covered by Simbol–X, and of the corresponding very large number of corresponding targets, Simbol–X is designed to offer two full years of scientific data taking, and to have the sufficient propellant resources to accommodate over 1000 different targets, with observation times ranging from ~ 20 ks for the brightest ones to 1 Ms or over for deep fields.

A relatively hard point of the design is the implementation of the collimator and sky baffle. In addition to its length, the metallic collimator should also not be a source of trouble for the systems in charge of the intersatellite links as well as those needed for the measurements of the relative positioning of the two spacecrafts. Depending on its final length, the collimator might be either implemented with the focal plane unit positioned on the spacecraft side facing the mirror, or in a configuration which puts the focal plane unit on the other side of the spacecraft, with the collimator going through the spacecraft structure, rendering the system more compact. The Figure 6 shows drawings of the mirror and detector spacecrafts corresponding to the first case, with a short collimator of ~ 1.5 m. The sky baffle on the mirror spacecraft, which has a diameter of 3 m here, can be seen together with the mechanical interface with the launcher.

The envisioned launch vehicle is a Soyuz with a Fregat upper stage. The fairing offers ample space for fitting the two spacecrafts, with two different options. One is to mate the two spacecrafts together, as a single composite spacecraft, for the launch operations up to the time when the composite reaches the operational orbit; the detector and mirror spacecrafts would then separate. The other option is to have the two spacecrafts as a dual launch, each of them reaching the operational orbit independently. This second option offers the advantage of simplifying all mechanical interfaces between the two spacecrafts; the Figure 6 drawings correspond to that case.

Detailed mass estimates have been done, including consumables for the observation program capabilities mentioned above, and the interfaces needed with the launcher. The total mass to be launched is around 2.2 tons, including the required margins at this level of study, for a launch capability of ~ 2.3 tons for the desired orbit.

## 5. CONCLUSIONS

The Simbol–X mission will provide an unprecedented sensitivity and angular resolution in the high energy domain, enabling to solve a number of outstanding questions in the non-thermal universe. Selected as a possible payload following the CNES call for ideas of a scientific mission for a formation flying demonstrator, it has undergone a detailed study which shows the feasibility of the mission within the given programmatic and technical constraints. Simbol–X is designed for a launch at the end of 2012, and for 2 years of scientific observations.

Simbol–X is expected to be further selected in fall 2005 for a competitive phase A, together with one other mission. As discussed in this paper, compared to the nominal initial baseline of paper I, it appears today possible to further improve the performances of Simbol–X, in particular for enlarging the field of view and increasing the effective area around 80 keV. This will be studied in depth during the phase A.

## ACKNOWLEDGMENTS

We are profoundly indebted to all the people who are actively participating to the elaboration of the Simbol–X scientific proposal, as well as to the hardware development, and who are too numerous to be cited here. We also warmly acknowledge the PASO team of CNES, led by Paul Duchon and Rodolphe Cledassou, for the thorough Simbol–X mission study they have performed, as well as for allowing publication of material from this study in this paper.